\documentstyle[prd,preprint,aps]{revtex}
\begin{document}
\draft
\title{Optimal Guessing Strategies in a Quantum Card Game}
\author{C.-L. Chou\footnote{choucl@cycu.edu.tw} and L. Y. Hsu\footnote
{cdhu6@phys.ntu.edu.tw}}
\address{$^*$Department of Physics, Chung-Yuan Christian University,
Taoyuan, Taiwan 32023}
\address{$^\dag$Department of Physics, National Taiwan University,
Taipei, Taiwan 10673}
\date{\today}
\maketitle
\begin{abstract}
Three different quantum cards which are non-orthogonal quantum
bits are sent to two different players, Alice and Bob, randomly.
Alice receives one of the three cards, and Bob receives the
remaining two cards. We find that Bob could know better than Alice
does on guessing Alice's card, no matter what Bob chooses to
measure his two cards collectively or separately. We also find
that Bob's best strategy for guessing Alice's card is to measure
his two cards collectively.
\end{abstract}
\pacs{ }
\section{Introduction}
It has been questioned by Peres and Wootters\cite{wootters} some
time ago that whether more information can be extracted from a
composite quantum system by performing collective measurements on
the system as a whole. Several studies addressed the same question
and showed that collective measurements usually provide more
information than the measurements on the individual
subsystems\cite{massar,scudo,gisin,bagan}. Quantum entanglement is
believed to be responsible for this. In this paper, we want to
address another question for composite quantum systems. Suppose a
quantum system is composed of two entangled subsystems A and B.
Which one of the two subsystem provides more information than the
other subsystem? To give more insight to the question, we
investigate the following quantum card game.

The game that we consider has two players, Alice and Bob, and one
card dealer. The dealer holds three quantum cards which are
non-orthogonal spin-$1/2$ particles $|\psi_{1}>=(1,0)$,
$|\psi_{2}>=(1/2, \sqrt{3}/2)$ and $|\psi_{3}>=(-1/2,\sqrt{3}/2)$,
respectively. In the beginning of the game, the dealer shuffles
the cards randomly and gives Alice one of the cards. Bob picks the
remaining two cards from the dealer. Both players then make their
own guesses on what card is in Alice's hand. After making their
guesses, the dealer will check their answers and decide who is the
winner of the game. Based upon the game description, we can view
the dealer's cards as a composite quantum system $\rho$ and both
Alice's card and Bob's cards as two subsystems, $\rho_A$ and
$\rho_B$, of the composite system.
\begin{eqnarray}
\rho&=&{1\over
6}\{|\psi_1><\psi_1|_A\otimes|\psi_2\psi_3><\psi_2\psi_3|_B
+|\psi_1><\psi_1|_A\otimes|\psi_3\psi_2><\psi_3\psi_2|_B\nonumber
\\ &+& |\psi_2><\psi_2|_A\otimes|\psi_1\psi_3><\psi_1\psi_3|_B
+|\psi_2><\psi_2|_A \otimes|\psi_3\psi_1><\psi_3\psi_1| \nonumber
\\&+& |\psi_3><\psi_3|_A\otimes|\psi_1\psi_2><\psi_1\psi_2|_B
 +|\psi_3><\psi_3|_A \otimes |\psi_2\psi_1><\psi_2\psi_1|_B \} \\
 \rho_A&\equiv&Tr_{B}(\rho)={1 \over 3}
(|\psi_1><\psi_1|_A+|\psi_2><\psi_2|_A+|\psi_3><\psi_3|_A)
\nonumber \\ &=& diag (1/2, 1/2) \\
 \rho_B&\equiv&Tr_{A}(\rho) \nonumber \\
 &=&{1\over 6}\{|\psi_1\psi_2><\psi_1\psi_2|_B+|\psi_2\psi_1>
<\psi_2\psi_1|_B+|\psi_1\psi_3><\psi_1\psi_3|_B \nonumber
\\&+&|\psi_3\psi_1><\psi_3\psi_1|_B+|\psi_2\psi_3>
<\psi_2\psi_3|_B+|\psi_3\psi_2><\psi_3\psi_2|_B \}
\end{eqnarray}
\noindent In the classical world, all the cards are orthogonal to
each other and thus totally distinguishable. Alice knows exactly
what her card is by simply looking at the card she has. On the
other hand, Bob also concludes on what Alice has in hand by simply
looking at his two cards. Both players always agree with each
other and thus there is no possibilities of cheating for both
Alice and Bob. Therefore, it is not likely for them to gamble on
what card is in Alice's hand because of the lack of uncertainty in
the game. That is, both Alice and Bob posses the same amount of
information on Alice's card.

In the quantum world, the non-orthogonal nature of the cards take
parts in this game. The quantum cards are no longer totally
distinguishable. Therefore the game becomes uncertain and the
gambling between Alice and Bob is possible. Alice no longer
determines what she has in hand with $100$ percent certainty and
so does Bob. For instance, if Alice performs Stern-Gerlach
measurement on her card along the direction
$|\phi_{\alpha}>=(\cos{\alpha},\sin{\alpha})$, she would obtain
the spin-up or the spin-down result with the following
probabilities:
\begin{eqnarray}
P(up)&=&Tr\{|\phi_{\alpha}><\phi_{\alpha}|\rho_{A}\} \nonumber \\
P(down)&=&1-P(up) \label{eqn:Aliceprob1}
\end{eqnarray}
\noindent In the game considered, Alice always measures the
spin-up or spin-down outcomes with equal probabilities,
$P(up)=P(down)=1/2$, no matter what the measuring direction
$|\phi_\alpha>$ is. As shown in Table. \ref{tbl:AcondProb}, when
the outcome $r$ is obtained after the measurement, Alice knows $a$
$posteriori$ the probability $P(i|r)$ for preparation $|\psi_i>$.
She simply cannot determine what her card is with $100$ percent
certainty.

\section{The optimal strategies for Alice}
Alice has one quantum card in hand. The card is actually a
spin-$1/2$ particle which is described by $\rho_A$. As seen from
the previous section, all $|\psi_1>$, $|\psi_2>$ and $|\psi_3>$
are spin-up directions on the $ZX$ plane and equally separated.
Thus the density operator $\rho_A$ possesses a $Z_3$ symmetry
which leaves $\rho_A$ invariant under the rotation about Y axis
$|\psi_i> \rightarrow R_y(2\pi/3)|\psi_i>$. Therefore Alice may
choose to measure her card along $|\phi_\alpha> = (\cos\alpha,
\sin\alpha)$ direction $(-\pi/6 \leq \alpha \leq \pi/6)$ on the
$ZX$ plane without lose of generality.

In the $\alpha$ range that we consider, the conditional
probability $P(1|r)$ is the largest one among all probabilities
$P(i|r)$ when the measurement outcome $r$ is spin-up. Thus Alice
should guess $|\psi_1>$ on her card when she measures a spin-up
outcome. On the other hand when the outcome is spin-down, Alice
may adopt three strategies. The first strategy is to guess
$|\psi_3>$ for $0 < \alpha \leq \pi/6$. It thus leads to the
successful guessing probability for Alice:
\begin{equation}
P(success)={1\over 3}\{\cos^2\alpha + \sin^2(\alpha+{\pi \over
3})\}. \label{eqn:ASuccProb}
\end{equation}
\noindent It is easily found that when $\alpha={\pi / 12}$, the
first strategy has an optimal value
$P(success)={(2+\sqrt{3})/6}\approx 0.622$. The second strategy is
guessing $|\psi_2>$ for $-\pi/6 \leq \alpha <0$. This one has the
same optimal guessing probability as the first strategy at $\alpha
= {-\pi /12}$. The third strategy is to randomly guess $|\psi_2>$
or $|\psi_3>$ with equal probability for $\alpha=0$. This strategy
is not optimal simply because it has a smaller successful
probability of guessing.

It is noted that the optimal probability can also be obtained by
minimizing Shannon entropy. For example, Alice choose the first
strategy to guess her card in hand. When the measurement outcome
is spin-up, Alice will guess her card successfully with
probability $(2\cos^2\alpha) /3$, and make a wrong guess with
probability $1-(2\cos^2\alpha)/3$. When the measurement outcome is
spin-down, she guesses the card successfully with probability
$2\sin^2(\alpha+\pi/3) /3$ and guesses it wrong with probability
$1- 2\sin^2(\alpha+\pi/3) /3$. Thus the Shannon entropy $S$ for
the "success-failure" binary information is:

\begin{eqnarray}
S&=&-\sum_{r}P(r)\{P(success|r) Log_2P(success|r)+P(failure|r)
Log_2P(failure|r)\} \nonumber \\&=&-({1\over 3}\cos^2\alpha
)Log_2({2\over 3}\cos^2\alpha)-({1\over 2}-{1\over
3}\cos^2\alpha)Log_2(1-{2\over 3}\cos^2\alpha) \nonumber \\
&-&{1\over 3}\sin^2(\alpha+{\pi \over 3})Log_2({2\over
3}\sin^2(\alpha+{\pi \over 3}))-({1\over 2}-{1\over
3}\sin^2(\alpha+{\pi \over 3}))Log_2(1-{2\over
3}\sin^2(\alpha+{\pi \over 3}))
\end{eqnarray}
\noindent The minimum of the entropy $S$ occurs at $\alpha=
\pi/12$ for $0<\alpha\leq \pi/6$.

\section{The optimal strategies for Bob}
Bob receives two cards from the dealer. Before measuring his
cards, Bob knows that there are six possible combinations for the
his cards. They are ${|A>} {\equiv} {|\psi_1\psi_2>}$,
$|A^{'}>\equiv |\psi_2\psi_1>$, $|B> \equiv |\psi_2\psi_3>$,
$|B^{'}> \equiv |\psi_3\psi_2>$, $|C> \equiv |\psi_3\psi_1>$ and
$|C^{'}> \equiv |\psi_1\psi_3>$. These six states are not
orthogonal states. That means Bob is no way of distinguishing them
with $100$ percent certainty. However, Bob does not really need to
know exactly what he has in hand to infer Alice's card. For
example, Bob will infer that Alice's card is $|\psi_3>$ if he
thinks he has $|A>$ or $|A^{'}>$ in hand. If Bob has $|B>$ or
$|B^{'}>$, he will infer $|\psi_1>$ for Alice's card. Similarly if
Bob has $|C>$ or $|C^{'}>$, he will think that Alice has the
$|\psi_2>$ card in hand.

To make a reasonable guess, Bob needs to measure his cards before
guessing. There are two kinds of the measuring methods: $(1)$
Measure the two cards on by one, or $(2)$ Measure them
collectively. For the first method, Bob will measure his first
card correctly with the optimal probability $P_1=(2+\sqrt{3})/6$.
The second card then can be measured correctly with the optimal
probability $P_2=(2+\sqrt{3})/4$. Thus Bob can know his cards
exactly with the optimal probability $P_{12}$:
\begin{equation}
P_{12}=P_1\times P_2 = {7+ 4 \sqrt{3} \over 24}\approx 0.5803
\end{equation}
\noindent However, except for measuring his cards successfully Bob
can also infer the same card for Alice by making two consecutive
incorrect guesses on his cards. For example, assume that Bob does
receive $|\psi_2\psi_3>$ cards from dealer. If the measurement
result is $|\psi_3\psi_2>$, he still infer the same result
$|\psi_1>$ for Alice. For this case, the probability of a
successful guess is $P_{21}$:
\begin{equation}
P_{21}=(1-{2+\sqrt{3} \over 6}) \times {1 \over 2} \times {1\over
2}={4-\sqrt{3} \over 24}
\end{equation}
\noindent Therefore, the optimal probability that Bob will infer
the correct card for Alice is $P_{Bob}(separate)$:
\begin{equation}
P_{Bob}(separate)=P_{12}+P_{21}={11+3\sqrt{3} \over 24} \approx
0.6748
> 0.6220
\end{equation}
\noindent It is obviously that by measuring his cards one by one,
Bob has a higher optimal guessing probability than Alice does on
guessing Alice's card.

The other measuring method is doing combined measurement on Bob's
cards. Since the dimension of the Hilbert space for Bob's cards is
$4$, Bob needs to choose a suitable orthonormal basis $\{|\phi_1>,
|\phi_2>, |\phi_3>, |\phi_4>\}$ in the Hilbert space for his
measurement.
\begin{eqnarray}
|\phi_1>&=&a_1|A_1>+a_2|A_2>+a_3|A_3>+a_4|A_4> \\
|\phi_2>&=&b_1|B_1>+b_2|B_2>+b_3|B_3>+b_4|B_4> \\
|\phi_3>&=&c_1|C_1>+c_2|C_2>+c_3|C_3>+c_4|C_4>
\end{eqnarray}
\noindent where the base vectors $|\phi_1>$, $|\phi_2>$, and
$|\phi_3>$ are written in terms of other orthonormal bases
$\{|A_i>\}$, $\{|B_i>\}$, and $\{|C_i>\}$ in the Hilbert space,
respectively.
\begin{eqnarray}
|A_1>&\equiv&\sqrt{2 \over 5}(|A>+|A'>), \hspace{0.8cm} |A_2>
\equiv \sqrt{2 \over 3}(|A>-|A'>) \nonumber \\ |A_3>&\equiv&
\sqrt{2 \over 3}(|\psi_1,\psi_1>-|\psi_2,\psi_2>),\nonumber \\
|A_4> &\equiv& {1\over 3}\sqrt{2\over
5}(|\psi_1,\psi_1>+|\psi_2,\psi_2>+4|\psi_3,\psi_3>) \nonumber \\
|B_1>&\equiv&\sqrt{2 \over 5}(|B>+|B'>), \hspace{0.8cm} |B_2>
\equiv \sqrt{2 \over 3}(|B>-|B'>) \nonumber \\ |B_3>&\equiv&
\sqrt{2 \over 3}(|\psi_2,\psi_2>-|\psi_3,\psi_3>),\nonumber \\
|B_4> &\equiv& {1\over 3}\sqrt{2\over
5}(|\psi_2,\psi_2>+|\psi_3,\psi_3>+4|\psi_1,\psi_1>) \nonumber \\
|C_1>&\equiv&\sqrt{2 \over 5}(|C>+|C'>), \hspace{0.8cm} |C_2>
\equiv \sqrt{2 \over 3}(|C>-|C'>) \nonumber \\ |C_3>&\equiv&
\sqrt{2 \over 3}(|\psi_3,\psi_3>-|\psi_1,\psi_1>),\nonumber \\
|C_4> &\equiv& {1\over 3}\sqrt{2\over
5}(|\psi_3,\psi_3>+|\psi_1,\psi_1>+4|\psi_2,\psi_2>)
\end{eqnarray}
\noindent The unknown coefficients $a_i$, $b_i$, and $c_i$ should
be determined by the optimal guessing strategies and are assumed
to be real numbers for simplicity. They satisfy the following
orthonormal conditions:
\begin{eqnarray}
\sum^4_{k=1}a_k^2 &=&1 \nonumber \\ \sum^4_{k=1}b_k^2&=&1
\nonumber
\\ \sum^4_{k=1}c_k^2&=&1 \nonumber \\
<\phi_i|\phi_j>&=&\delta_{ij}.
\end{eqnarray}
\noindent It is noted that the base vector $|\phi_4>$ can be
determined once the other vectors $|\phi_i>$ are determined. Any
measurement $M$ along the $|\phi_4>$ direction can be related to
the measurements along other directions.
\begin{equation}
Tr(|\phi_4><\phi_4|M)=Tr(M)-\sum^3_{k=1}Tr(|\phi_k><\phi_k|M)
\end{equation}
\noindent With the orthonormal measuring basis, Bob makes the
following guessing strategy: $(a)$ Guess $|\psi_3>$, $|\psi_1>$,
or $|\psi_2>$ for Alice's card when the measurement outcome is
$|\phi_1>$, $|\phi_2>$, or $|\phi_3>$, respectively. $(b)$ When
the outcome is $|\phi_4>$, Bob has three choices in general:(I)
Guess one of the three cards for Alice. For example, guess
$|\psi_3>$ for Alice. (II) Guess Alice's card randomly from two of
the three possible cards. For example, randomly guess $|\psi_3>$
or $|\psi_1>$. (III) Make a random guess from the three cards. In
general, different guessing choices for the outcome $|\phi_4>$ may
lead to different optimal guessing probabilities. However,
numerical study shows that all these three choices have the same
optimal guessing probabilities which correspond to the same
measuring basis. Therefore, we will discuss only the third
guessing choices in the paper.

With the guessing strategy, Bob knows that the probability for a
successful guessing on Alice's card is $P_{Bob}(combined)$:
\begin{eqnarray}
P_{Bob}(combined)&=&{1\over
6}\{<A|\phi_1>^2+<A'|\phi_1>^2+<B|\phi_2>^2+<B'|\phi_2>^2
\nonumber \\ &+&<C|\phi_3>^2+<C'|\phi_3>^2+{1\over
3}(<A|\phi_4>^2+<A'|\phi_4>^2 \nonumber \\&+&
<B|\phi_4>^2+<B'|\phi_4>^2+<C|\phi_4>^2+<C'|\phi_4>^2 )\}
\nonumber \\&=&{ 1\over 3}+{2\over 15}(a_1^2+b_1^2+c_1^2)-{1\over
12}(a_3^2+b_3^2+c_3^2) \nonumber \\ &-&{1\over
20}(a_4^2+b_4^2+c_4^2) +{1\over 30}(a_1a_4 +b_1b_4 -c_1c_4)
\end{eqnarray}
\noindent Numerical study shows that when
$a_1=b_1=c_1=(4+\sqrt{2})/\sqrt{30}$, $a_2=b_2=c_2=0$,
$a_3=b_3=c_3=0$, and $a_4=b_4=-c_4=(2-\sqrt{2})/\sqrt{15}$, the
successful probability $P_{Bob}(combined)$ has a maximum value
$P_{Bob}(combined)=(3+\sqrt{2})/6\approx 0.7357$. That means, the
best guessing strategy for Bob to know Alice's card is doing
combined measurement on his cards.

\section{Conclusion}
In this paper, we study the quantum card game of two players,
Alice and Bob, who receive their cards from a card dealer. The
dealer shuffles three quantum cards randomly and then sends Alice
one of the cards. Bob then receives the remaining two cards from
the dealer. In the classical world, the three cards are actually
orthogonal to each others and thus are totally distinguishable.
Therefore, Alice knows her card with $100$ percent confidence, and
Bob also knows Alice's card with $100$ percent confidence by
simply looking at his two cards.

In the quantum world, the three cards can be viewed as three
quantum bits and are in general non-orthogonal to each other. The
non-orthogonality forces both Alice and Bob to do measurements on
their cards and then make their guesses. Our study shows that the
best chance for Alice to know her card is
$P_{Alice}=(2+\sqrt{3})/6\approx 0.6220$, a probability which is
nearly doubled than a random guess $P=1/3$. On the other hand, Bob
can choose either to measure his cards one by one or to measure
them collectively. Both ways of measuring all give Bob the optimal
guessing probabilities $P_{Bob}(separate)=(11+3\sqrt{3})/24\approx
0.6748$ and $P_{Bob}(combined)=(3+\sqrt{2})/6\approx 0.7357$ which
are larger than Alice's optimal guessing probability. That is, Bob
has a higher winning probability than Alice does in the quantum
guessing game.
\subsection*{Acknowledgments}

This work was supported in part by National Science Council of
Taiwan NSC90-2112-M-033-011.


\begin{table}
\begin{tabular}{|c|c|c|c|} &\makebox[40mm]{$P(1|r)$}&
\makebox[40mm]{$P(2|r)$}&\makebox[40mm]{$P(3|r)$} \\ \hline
spin-up & ${2 \over 3} \cos^2\theta$&${2 \over
3}\cos^2(\theta-{\pi \over 3})$&${2 \over 3} \cos^2(\theta+{\pi
\over 3})$
\\ \hline spin-down & ${2 \over 3}
\sin^2\theta$&${2 \over 3}\sin^2(\theta-{\pi \over 3})$&${2 \over
3} \sin^2(\theta+{\pi \over 3})$ \\
\end{tabular}
\vspace{0.3cm} \caption{Conditional probabilities for Alice's
measurement outcomes} \label{tbl:AcondProb}
\end{table}

\begin{references}
\bibitem{wootters}A. Peres and W. K. Wootters, {\it Phys. Rev.
Lett.} {\bf 66}(1991)1119.
\bibitem{massar}S. Massar and S. Popescu, {\it Phys. Rev. Lett.} {\bf
74}(1995)1259.
\bibitem{scudo}A. Peres and P. F. Scudo, {\it Phys. Rev. Lett.} {\bf
86}(2001)4160.
\bibitem{gisin}N. Gisin and S. Popescu, {\it Phys. Rev. Lett.} {\bf
83}(1999)432.
\bibitem{bagan}E. Bagan, M. Baig, A. Brey, and R. Munoz-Tapia,
{\it Phys. Rev. A} {\bf 63}(2001)052309.
\end{references}
\end{document}